\documentclass[twocolumn, 10pt, floatfix, superscriptaddress, booktabs]{revtex4}

\usepackage{amsmath}
\usepackage{amssymb}
\usepackage{amsfonts}   
\usepackage{graphicx}   
\usepackage{verbatim}   
\usepackage{color}      
\usepackage{subfigure}
\usepackage{pgfplotstable}
\usepackage{wrapfig}
\usepackage{hhline}

\usepackage{tikz}
\usetikzlibrary{positioning}
\definecolor{myRed}{RGB}{229,25,50}
\definecolor{myBlue}{RGB}{25,178,255}
\definecolor{myGreen}{RGB}{50,255,0}

\begin{document}
\title{Avalanche Burst Invasion Percolation Model: Emergent Scale Invariance on a Pseudo-Critical System}
\date{\today}

\author{Ronaldo Ortez}
\email{raortez@ucdavis.edu}
\affiliation{Department of Physics, One Shields Ave., University of California, Davis, CA 95616, United States}
\author{John B. Rundle}
\email{rundle@ucdavis.edu}
\affiliation{Department of Physics, One Shields Ave., University of California, Davis, CA 95616, United States}
\affiliation{Department of Geology, One Shields Ave., University of California, Davis, CA 95616, United States}
\affiliation{Santa Fe Institute, Santa Fe, NM 87501, United States}

\date{\today}
\begin{abstract}
	Given the large variety and number of systems displaying scale invariant characteristics, it is becoming increasingly important to understand their fundamental and universal elements. Much work has attempted to apply second order phase transition mechanics due to the emergent scale invariance at the critical point. However for many systems, notions of phases and critical points are both artificial and cumbersome. We characterize the critical features  of the avalanche-burst invasion percolation(AIP) model since it exists as a hybrid critical system (of which many self-organized critical systems may fall under). We find behavior strongly representative of critical systems, namely, the presence of a critical Fisher type distribution, $n_s(\tau, \sigma)$, but other essential features absent such as an order parameter and to a lesser degree hyperscaling. This suggests that we do not need a full phase transition description in order to observe scale invariant behavior, and provides a pathway for more suitable descriptions. 
\end{abstract}

\maketitle

\section{Introduction}
With the advent of complexity sciences there has been a steady discovery of systems which display behavior characterized by power-law (fractional) statistics. Such systems fall broadly under the umbrella of fractal, chaos, and critical theory with many interesting links spanning these systems. As a cornerstone example, many authors have noted, in classical, tectonic seismicity, essentially all of the statistical properties are described by a power-law relations: magnitude frequency distribution(G-R magnitude scale)\cite{rundle1989derivation}, temporal aftershock clustering(Omori aftershock law)\cite{shaw1993generalized}, and the two-point correlation distribution\cite{kagan2007earthquake}. Similarly, induced seismicity also follows these statistical measures where in hydraulic fracturing\cite{ebrahimi2010invasion} and geothermal injection\cite{henderson1999fractal} studies show power-laws describing both the magnitude-frequency distribution and two-point correlation function. Coupled with fractal fault and fracture properties, seismicity offers a plethora of tantalizing critical characteristics.

One of the key insights from the study of critical theory and second order phase transitions is that systems described by power-laws manifest a particular symmetry known as scale-invariance. Scale invariance suggests that there is no fundamental scale associated with the system. In many cases, this is observed to correspond to a self-similarity across many different length scales, and an essential question is how these properties might arise naturally and in all their varied forms. In second order phase transitions, scale-invariance emerges when behavior becomes dominated by the system's inherent random fluctuations which are allowed to grow to all scales of the system\cite{stanley1971phase}. This boundary between the order, dictated by the small scale physics governing the properties of distinct phases, and the emergence of critical fluctuations, is precisely described by critical theory. Fundamentally, this transition is stated as the finite-size scaling hypothesis which posits that the divergence of the system’s correlation length is the mechanism by which realizations of critical fluctuations are allowed to grow to any accessible scale\cite{ma2018modern}. As a consequence, notions of universality emerged to form distinct universality classes, where universality classes describe power-law statistics shared by many distinct systems, and highlights their shared stochastic nature. These insights applied to complex systems nurtured a general approach and led to a tremendous amount of work in the last 50 years.


Percolation theory, being the simplest example of a fully featured critical theory, provided the basis for understanding the essential elements which served to define distinct universality classes. In particular, it was found that the thermal second order phase transition of the 2-D Ising model could be mapped to the much simpler percolation problem \cite{murata1979hamiltonian,coniglio1980clusters,klein2007structure,reynolds1977ghost}. As variations to percolation served to define distinct universality classes, authors naturally sought to find the corresponding physical systems matching the behavior. 

Perhaps equally important, much of critical behavior can be neatly represented by the critical clusters describing the system's emergent connectivity near the critical point. This can be nicely represented by Fisher's formulation of cluster size distribution parameterized by two exponents, $\sigma, \tau$,
\begin{equation}
n_s\left(T\right) = s^{-\tau} f\left[\left(T-T_c\right)s^\sigma\right]
\label{eq:clusterDist}
\end{equation} 
The existence and characterization of such a quantity is a good starting point in describing the critical behavior that exists in a system. 

Wilkenson proposed IP as a more dynamic variant to RP aimed at describing the invasion of fluid within a sediment guided by the path corresponding to that of least resistance through the sediment. This model deviates fundamentally from RP, since it only describes a growth mechanism that "self-organizes" to produce a scale invariant cluster. 
Many aspects of IP have been studied, including various fractal dimensions \cite{willemsen1984investigation,sheppard1999invasion}, its cluster characteristics \cite{ebrahimi2008shape}, effects of long-range correlations \cite{knackstedt2000invasion}, and its dynamics in correlated porous media \cite{Vidales_1996}.

The aim of this paper is two-fold. First, we describe how critical theory manifests itself in IP (and since IP is a prime example of self-organized criticality(SOC), perhaps SOC at large). This can only be done in part as IP alone lacks the requisite critical cluster distribution. Second, we propose a modest extension with avalanche-burst dynamics to make the analysis more precise. Briefly, we introduced the notion of a connected burst\cite{ortez2021universality} defined as the sequential invasion of sites/bonds below some threshold strength, and found that there existed a critical threshold value which produced a scale invariant burst distribution characterized by scaling exponent $\tau$. Away from the critical threshold, we observe the existence of a cutoff cluster characterizing the $\sigma$ dependence. This produces a distinct Fisher type distribution given by Eq.\ref{eq:clusterDist} from which we can characterize the universality class of our model.

\section{Boundary Conditions}
\label{section:boundaryConditions}
Because the exponents that define a particular universality class are taken to be in the infinite lattice limit, we imposed periodic edge boundary (PEB) conditions in this paper. In a previous study, we characterized some of the essential network properties of our AIP model \cite{ortez2021universality} with free edge boundary(FEB) conditions along both growth axes, and achieved the infinite lattice limit by extrapolating the finite size scaling of the exponents to the infinite limit. However, a similar extrapolation of finite size effects for all quantities becomes extremely cumbersome, and we, like other authors \cite{kenna2017universal, heermann1980influence}, have found how non-trivially finite size effects can differ from the infinite lattice limit.


As scale invariance is more manifest with (PEB) conditions, it reduces finite size corrections required to obtain infinite size limits for many of the system's scaling exponents \cite{ziff1997universality}. For this reason and because we sought to characterize many more of the scaling exponents, we chose to implement (PEB) conditions in this study.  

\begin{figure*}[t]
	\includegraphics[width=1.0\textwidth]{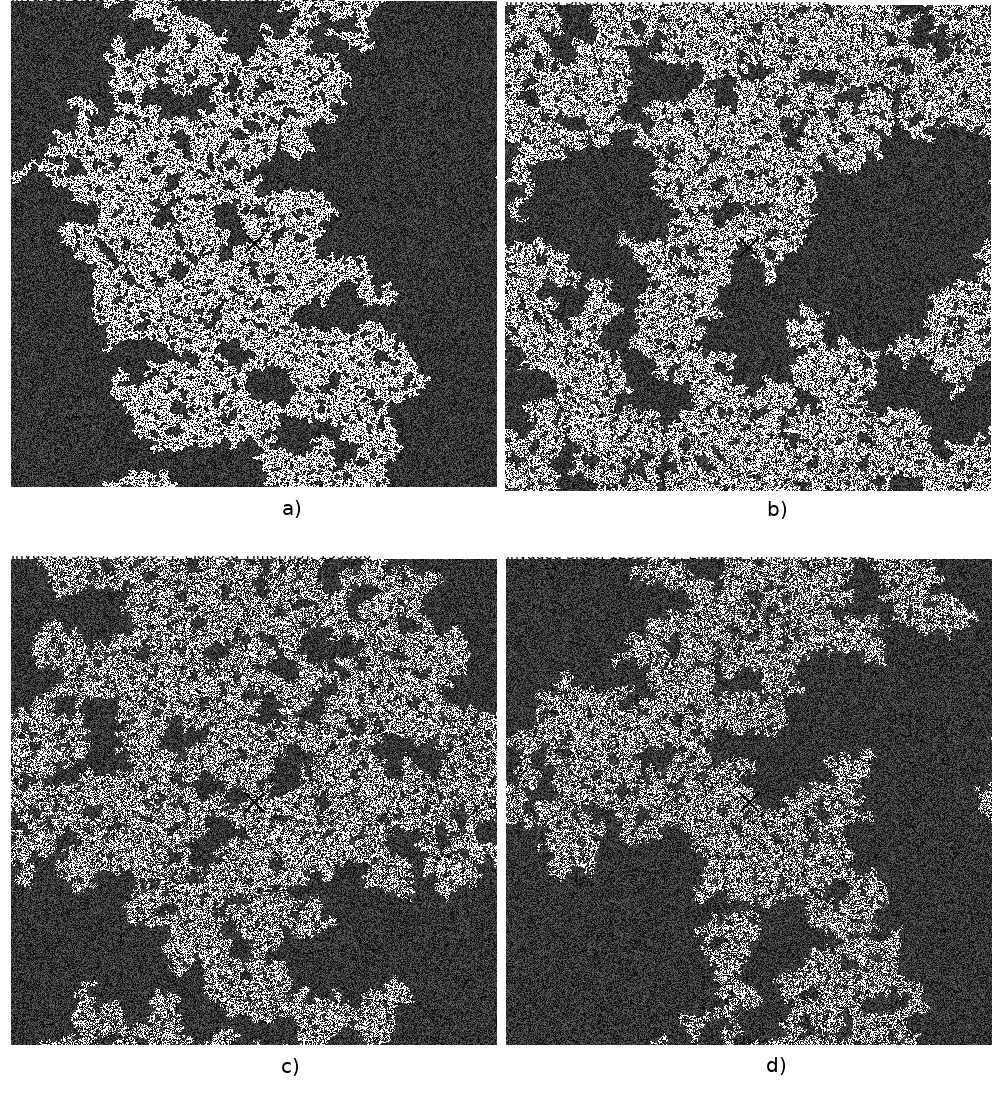}
	\caption{\footnotesize SIP algorithm with periodic boundary conditions (PBC)with four different lattice sizes corresponding to: a)500x500 b)1000x1000 c)5000x5000 d)10000x10000. }
	\label{fig:SIPScaleInv}
\end{figure*}

One of the primary challenges with implementing PEB conditions is coming up with a suitable stop condition. Most implementations of IP preserve some free edge to trigger a stop condition and make the others periodic\cite{wilkinson1983invasion,wagner1997fragmentation,knackstedt2002nonuniversality,ebrahimi2008shape}, but we chose to implement fully periodic boundaries following an implementation similar to \cite{watanabe1995percolation}. Because IP is isotropic, we expect the network to be manifestly self-similar, and we manufactured the stop condition by tracking cluster growth from the initial central seed towards and around a boundary. A stop condition is satisfied when a cluster wraps around the boundary and shares a perimeter site with a cluster of different boundary. For example, if the central cluster first reached the left boundary, then it now belongs to the left boundary cluster and is distinguished from sites growing from the right lattice boundary; these sites belong to the right boundary cluster. The central cluster can belong to multiple boundary clusters and therefore activate multiple opposite boundary clusters which could trigger the stop condition. This stop condition nicely preserves self-similarity on all scales up to the lattice size. See Fig \ref{fig:SIPScaleInv} for cluster examples where the lattice size is very difficult to decipher simply by inspection. 

\begin{figure*}[t]
	\includegraphics[width=1.0\textwidth]{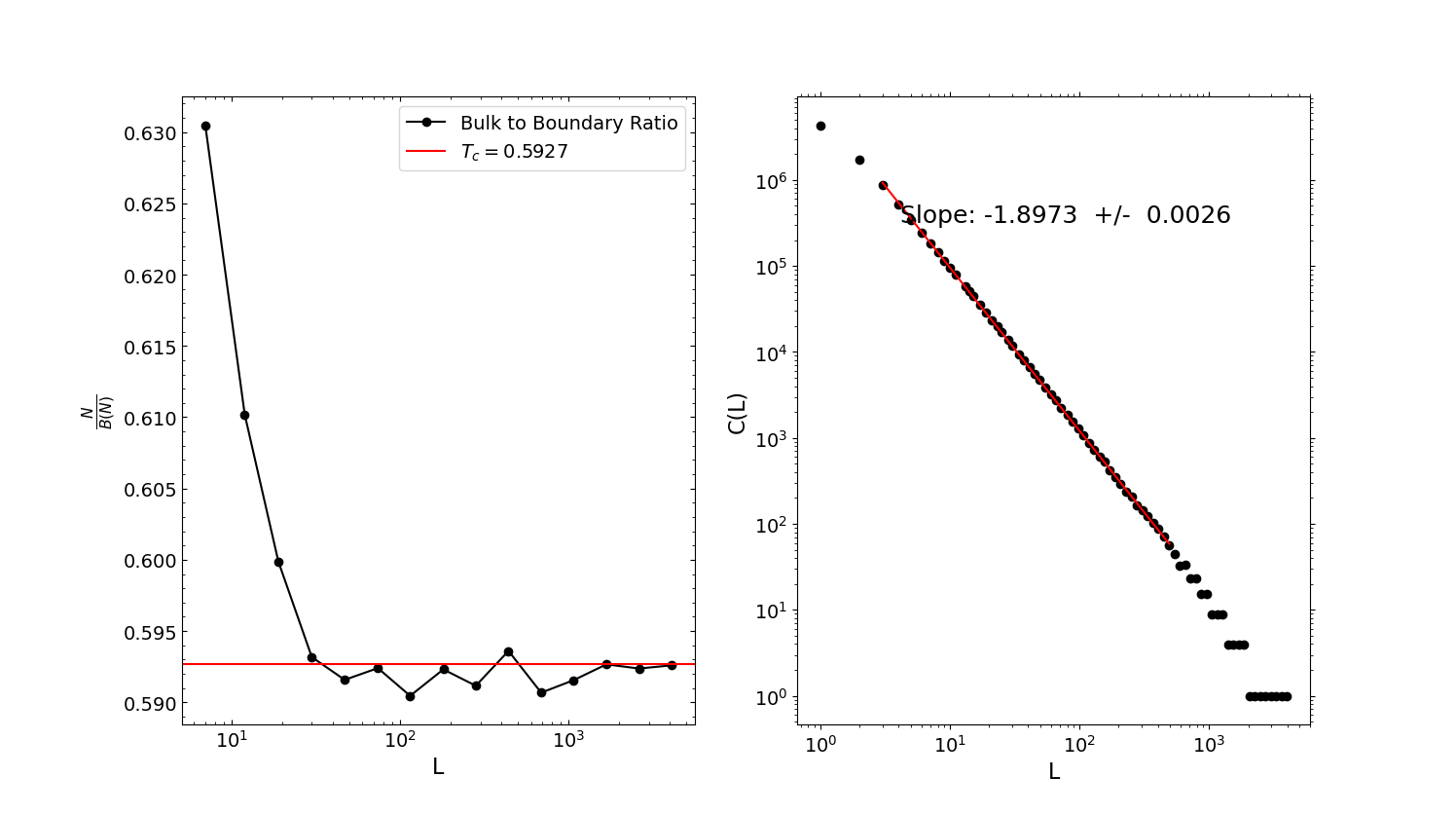}
	\caption{\footnotesize SIP algorithm with periodic boundary conditions (PBC). (Left) We show the bulk to boundary ratio of the invaded cluster as a function lattice size. This macroscopic geometric measure quickly approaches its critical value as lattice size increases. (Right) We compute fractal scaling exponent, $D_f$, using the boxing counting technique for clusters grown in a $4096x4096$ lattice.  }
	\label{fig:PBCscaleInv}
\end{figure*}

We can also refer to quantitative measures by observing the existence of scale symmetry relation $M \sim L^{D_f}$ where $D_f$ is the typical percolation mass scaling exponent, $D_f \approx 1.89$. In the right plot of Fig\ref{fig:PBCscaleInv}, we compute $D_f$ using the box counting technique on an ensemble of $100$ lattices with size 4096x4096. With this technique, the box count should scale with L according to $C(L) \sim L^{-D_f}$. We find $D_f=1.897 \pm 0.003$ which is very close to the expected infinite lattice limit without requiring any corrections.  

In the next section we introduce the notion of the bulk to boundary ratio which is a geometric property indicative of a critical cluster. We can also compute the bulk to boundary ratio for clusters grown in different lattice sizes. We find the bulk to boundary ratio quickly settles down to that which is characteristic of a critical cluster. This is shown as the left plot of Fig.\ref{fig:PBCscaleInv}. This ensures that our PBC implementation does not alter the geometry of a critical cluster by prematurely stopping cluster growth.  

Finally, the type of boundary affects the burst distribution given by Eq.\ref{eq:clusterDist}. In our previous study\cite{ortez2021universality}, we determined the critical scaling of our burst distribition, $\tau-1\approx 1.52$, with FEB conditions. If each cluster is grown in isolation such that a cluster grows until reaching threshold $T$ and repeat to produce a distribution of clusters, then one obtains a value $\tau -1 \approx 1.0$. If instead we implements periodic boundary conditions in the lattice, we find a value of $\tau-1 \approx 1.585$. In higher dimensions one would expect this value to approach the mean-field limit of $\tau-1=1.5$. Variations in the cluster size distribution as a result of varying environment conditions was also found in \cite{araujo2005invasion}. The choice of boundary is central to the characterization of critical behavior. 

\section{Critical Threshold}
\label{section:critThresh}
One of the most important features of percolation is its relation to critical phenomena \cite{stephen1976percolation}. Before discussing the critical cluster distribution directly, it's first worth discussing the control parameter and its critical value. In RP the site occupation probability, $p$, serves as the control parameter for RP's second order phase transition. This phase transition is described by the emergence of global connectedness where the many isolated clusters existing in regime $p<p_c$ conglomerate into a single lattice spanning cluster for $p \sim p_c$. To better elicit the connection between RP and IP, its instructive to understand how RP's critical point manifests itself in IP. 

We begin by looking at the distribution of site strengths of the invaded cluster. In IP, all lattice sites are randomly assigned values from a uniform distribution in the range [0,1], but when looking at the distribution of the strengths of invaded sites, we find the selection of strengths to be a regular subset of assigned strengths. In particular, in the limit where the number of invaded sites,$N$, becomes infinite, the invaded strength distribution is described by a step function:

\begin{displaymath}
\lim_{N \to \infty} p(r) = \left\{
\begin{array}{lr}
k & 0 \ge r \ge r_{max} \\
0 & r > r_{max}
\end{array}
\right.
\end{displaymath}

where a random strength, $r$, has constant probability $k$, of being invaded up to some strength, $r_{max}$. These are related according to $1/k = r_{max}$, and its been shown that $r_{max}=p_c$ where $p_c$ is RP's critical occupation probability \cite{chayes1985stochastic}.

Without a threshold, a cluster grown by IP will grow indefinitely, and reproduce many of the characteristic exponents of RP's incipient infinite clusters (IIC), which is why IP is believed to reproduce the emergent incipient infinite cluster which defines RP's connected state \cite{jarai2003invasion}. A threshold in IP effectively acts in the same way as the occupation probability for RP. We use this property to distincguish the growth of multiple clusters into bursts by defining a burst to be the sequential invasion of sites of strength less than the threshold. The threshold serves to separate independent realizations of clusters grown by the IP method, and these independent realizations will naturally lead to a distribution of cluster sizes. However, if the threshold is $r_{max} \geq p_c$ then it becomes possible that a cluster will grow indefinitely. The statistics of this process near the critical point will yield independent realization of RP's emergent IIC with cluster realization existing on all scales.

An alternative notion for a burst could rely instead on a "bulk to boundary" ratio. The idea is that critical clusters are described by a characteristic ratio where near $p_c$, the ratio between the perimeter of filled and unfilled sites was 1. Using a similar strategy authors have argued that using a "bulk to boundary" ratio as a generalized way to determine the critical occupation probability \cite{mertens2017percolation}, determining the ratio analytically using,
\begin{equation}
\lim_{N \to \infty} \frac{N}{B(N)}=T_c
\label{eq:B2Bratio}
\end{equation} 
but this leads to $T_c \rightarrow p_c$ as before. Since both mechanism share the same critical point, its likely the cluster statistics are the same.   

\section{Critical Behavior}
\label{section:crit}
Since IP is believed to reproduce  the critical aspects of RP \cite{chayes1985stochastic}, we would expect the critical exponents of IP to be the same as those belonging to RP. When IP was initially proposed \cite{wilkinson1983invasion}, the critical behavior was believed to be described by the scaling of the acceptance profile, $B_1(n)$. This quantity, as described in the previous section, is a function of $p(r)$, the probability of invading a site/bond with strength, $r$. In the finite number $N$ limit, the acceptance profile is defined as,
\begin{equation} 
B_1(N)= \int_{0}^{p_c} [1-p_N(r)]dr
\end{equation}
where $N$ is the number of invaded sites. Near a strength matching percolation's critical value, $p_c$, it was believed that the region deviating from the $p_N(r)$ defining $B_1(N)$, was described by a power-law as function of the number of invaded sites, $N$. That is, near the critical value $B_1(N)$ would scale according to $ B_1(N) \sim N^{-1/\varDelta}$ with the gap exponent $ \varDelta = \beta+\gamma $, and $\gamma$ and $\beta$ are usual exponents from regular percolation. However, exponents $\gamma$ and $\beta$ are determined as functions of the first and second moments of RP's finite size cluster size distribution. This is an important observation as Fisher's critical droplet model \cite{fisher1967theory} distills criticality from the cluster distribution, and since there doesn't exist a cluster distribution native to IP, the source of IP's criticality becomes an open question. Perhaps this alone might not bring into question IP's criticality, if for example, it were also possible to discuss the characterization from the description of IP's order parameter, but as we will see, this too becomes problematic. 

The absence of a cluster distribution prohibited the authors of \cite{mertens2017percolation} from utilizing IP to directly obtain something analogous to the Fisher exponent, $\tau$, which in Fisher's formulation characterizes the distribution of non-interacting droplets near the critical point. It is only in trapping variants of IP where the notion of a defending fluid leads in a natural sense to a distribution of finite sized clusters. Still even in this case, studies of the scaling of these finite clusters show that they scale differently from RP finite clusters\cite{wilkinson1983invasion}. There is nonetheless valid reasons to suspect the appropriate relations between the criticality of RP and IP lies between IP's invasion cluster and RP's incipient infinite cluster \cite{jarai2003invasion}.
 
A full critical description identifies a distribution of fluctuations that dominate all scales of the system, and a characterization of the criticality is born out of the characterization of this distribution\cite{klein2007structure}. In the previous section, we established the notion of a control parameter with a critical value. We utilize the critical threshold to define distinct cluster realizations and thereby define a critical distribution of fluctuations as was done in our previous study\cite{ortez2021universality}. This allows a suitable definition of a Fisher distribution which follows the formulation,

\begin{equation}
n_s\left(T\right) = s^{-(\tau)} f\left[\left(T-T_c\right)s^{\sigma}\right]
\label{eq:clusterScaling}
\end{equation}

with cluster size becoming scale invariant for $T \rightarrow T_c$ and following power-law $n_s \sim s^{-(\tau-1)}$. We note that we follow the conventional definition of $n_s$ \cite{stauffer1994introduction}, and as such, $sn_s$ corresponds to the number of bursts of size $s$. This being the quantity we directly obtain, then has the shifted exponent, $\tau -1$ that we report.

Because we seek to understand IP's criticality from the perspective of traditional percolation, we will briefly describe the framework from which we will derive our results. A particularly useful formalism comes from previous work which shows the mapping between the Ising thermal second order phase transition to that of the percolation transition \cite{murata1979hamiltonian,coniglio1980clusters,klein2007structure,reynolds1977ghost}. Here one makes use of percolation's cluster distribution, $n_s$ to define an analogues generating function suitable for percolation from which the order parameter and other essential quantities may be derived.

Fundamentally, this mapping is possible because the thermal fluctuations of the order parameter(magnetization) experience long range correlations that allow large, macroscopic clusters of magnetization to emerge. The correlation function describing the correlation between sites behaves similarly in both systems, and the emergence of macroscopic magnetization clusters is therefore similar to the emergence of large connected percolation clusters. This behavior can be neatly summarized by characterizing the system's Fisher cluster distribution in addition to the behavior of its correlation length.
\begin{table*}[t]
	\begin{center}
		\setlength{\tabcolsep}{10pt}
		\begin{tabular}{l c c c } \hline \hline
			& AIP(This Paper) & RP\cite{stauffer1994introduction} & MFA\cite{landau2013statistical} \\ 
			\hline
			
			$D_f$ & $1.897(3)$ &$91/48 \approx 1.896$ & $2$   \\
			$D_{s}$ & $1.86(1)$& $91/48 \approx 1.896$ & $2$ \\
			$\tau$ &$1.594(9)$& $187/91 \approx 2.05$ & $3/2$   \\
			$\sigma$ & $0.41(2)$&$36/91\approx0.42$ & $1/2$ \\
			$\alpha$ & $0.10(5)$& $-2/3$& $-1$ \\
			$\beta$ &N/A &$5/36\approx.014$ & $1$ \\
			$\gamma$ &$0.971(5)$ &$43/18\approx2.4$ & $1$\\ 
			$\nu $ & $1.301(2)$& $4/3$& $1/2$\\
			\hline \hline
		\end{tabular}
		\caption{A comparison of critical exponents from the Avalanche-burst Invasion Percolation(AIP) present in this paper and regular percolation (RP) and mean field approximation values (MFA).}
		\label{table:critValues}
	\end{center}
\end{table*}

In much the same way that the free energy per spin behaves as the generating function for the Ising thermal critical transition, we define an analogous generating function from which we can derive RP's essential quantities. Beginning with the generating function,
\begin{equation}
G(n_s, h) = \int_{fc}ds \ \left\langle n_s \right\rangle e^{-hs} 
\label{eq:generatingFunc}
\end{equation} 
where one introduces a ghost field $h$, that allows cluster connectivity through this additional field $h$. For $h=0$ (the limit of the RP problem), the generating function becomes the mean number of finite sized clusters. A convenient feature of this formalism is that we can easily derive relations that define the order parameter, $P$, in terms of the first derivative of the generating function,
\begin{equation}
P = \frac{dG(n_s, h)}{dh}\bigg\rvert_{h=0} \propto \int_{fc}ds \ s\left\langle n_s\right\rangle
\end{equation}
which is proportional to the fraction of probability a site belongs to an infinite cluster. The order parameter is also proportional to the density of sites, thus, fluctuations in the order parameter may be more intuitively understood as fluctuations in local site density. 

Fluctuations in the order parameter can be calculated as the second derivative of the generating function. This would correspond to the second moment of the average cluster size distribution, $\left\langle n_s \right\rangle$, and coincidentally corresponds to the average cluster size, $\left\langle s \right\rangle$, with $s$ sites.
\begin{equation}
\left\langle s \right\rangle = \frac{d^2G(n_s, h)}{dh^2}\bigg\rvert_{h=0} \propto \int_{fc}ds \ s^2\left\langle n_s\right\rangle
\label{eq:meanCluster}
\end{equation}  

Next, the correlation length characterizes the spatial extent of system fluctuations. Traditionally, this is calculated by the pair connectivity function,
\begin{equation}
C(r, p) \sim \frac{1}{r^{d-2+\eta}} e^{-r/\xi(p)}
\end{equation}
which describes the likelihood two sites belong to the same cluster. It is dependent on the correlation length controlling when the sites begin to exhibit long range correlations such that for $p \rightarrow p_c$ the pair correlation function approaches power-law behavior described by $C(r,p_c)\sim r^{-(d-2+\eta)}$.
The emergence of long-ranged connectivity is usually empirically motivated and it is understood to be the consequence of the correlation length, $\xi$, diverging. The statement can actually be made more strongly and is known as the finite-size scaling hypothesis\cite{ma2018modern}. The claim is that all singular behavior is a result of a diverging correlation length. If $\xi$ becomes infinite, then the system lacks any intrinsic length scale and necessarily becomes scale invariant which is exactly the reason for the proliferation of power-law behavior describing all relevant variables. 

In RP, we can calculate the mean linear dimension of clusters, 
\begin{equation}
\left\langle \xi \right\rangle ^2 = \frac{1}{\left\langle s \right\rangle} \int ds \ \left\langle R_s \right\rangle^2s^2\left\langle n_s \right\rangle
\label{eq:corrLength}
\end{equation}
where $\left\langle R_s\right\rangle$ is the radius of gyration for a cluster with $s$ sites. This is calculated for each cluster of size s using,
\begin{equation}
R_s = \frac{\int dr (r -r_{cm})^2}{s}
\label{eq:gyr_radius}
\end{equation}
The average is performed over all clusters with $s$ sites. similarly $\left\langle s \right\rangle$ is the average cluster size consisting of $s$ sites. 

The critical behavior of the system corresponds to when the system is near its critical threshold, $p_c$. Here, all quantities become scale invariant and in R.P. are described by the reduced critical parameter,
\begin{equation}
\epsilon_{p} = \frac{p-p_c}{p_c} \ll 1
\end{equation}
Thus near the critical point we have, 
\begin{equation}
G \sim \epsilon_{p}^{1-\alpha}
\end{equation}
\begin{equation}
P \sim \epsilon_{p}^\beta
\end{equation}
\begin{equation}
\left\langle s \right\rangle \sim \epsilon_{p}^{-\gamma}
\label{eq:gammaScaling}
\end{equation}
\begin{equation}
\xi \sim \epsilon_{p}^{-\nu}
\end{equation}
Relation Eq.\ref{eq:numExcess} indicates that the number of bursts diverges as the critical parameter, $\epsilon_{T} \rightarrow 0 $. 

\section{AIP Criticality Results}
From relations \ref{eq:generatingFunc}-\ref{eq:corrLength} we can derive the scaling relations between burst distributions exponents and the characteristic quantities describing AIP criticality. In addition to illustrating the central role of the burst distribution, the relations will be used to derive exponent relations 17-20 and provide consistency checks for the results of the previous section. In order to establish these important relations, we must first independently determine the two exponents of the cluster distribution, $n_s(\tau, \sigma)$ and the scaling of burst size with its radius, $R_s$.

Following the technique in our previous study \cite{ortez2021universality}, except with PEB conditions we determine the power-law burst scaling $\tau-1 = 1.594\pm .0009$ for burst threshold, $T\sim T_c$. Next, following the characterization of \cite{stauffer1994introduction}, the $\sigma$ scaling exponent characterizes the behavior of the cutoff cluster size $s_\xi$ which itself defines the cutoff burst size such that bursts with size $s>s_\xi$ become exponentially suppressed. As is usual, we would like to determine the scaling of $s_\xi$ as we approach the critical point. In order to isolate the $\sigma$ behavior, we can eliminate the $\tau-1$ dependence by considering the ratio:
\begin{equation}
\frac{n_s(T)}{n_s(T_c)}=\exp[-z]
\label{eq:sigmaScaling]}
\end{equation}
where $z=(T_c-T)s^\sigma$. This leaves behind the exponential behavior of the cluster distribution. The two limits to consider is for $z \ll 1$ where the exponential becomes a constant, and in the large cluster limit where $z \gg 1$. In this limit, we observe the relation $s \gg (T_c -T)^{-1/
\sigma}$. Thus, $s_\xi(T) = (T_c -T)^{-1/\sigma}$ behaves as the exponential cutoff cluster size for the cluster size distribution. By determining the scaling of $s_\xi(T)$ with $T$ we can determine exponent $\sigma$, which for our cluster distribution is found to be $\sigma = 0.41 \pm 0.02$.

\begin{figure} 
	\centering
	\includegraphics[width=0.5\textwidth]{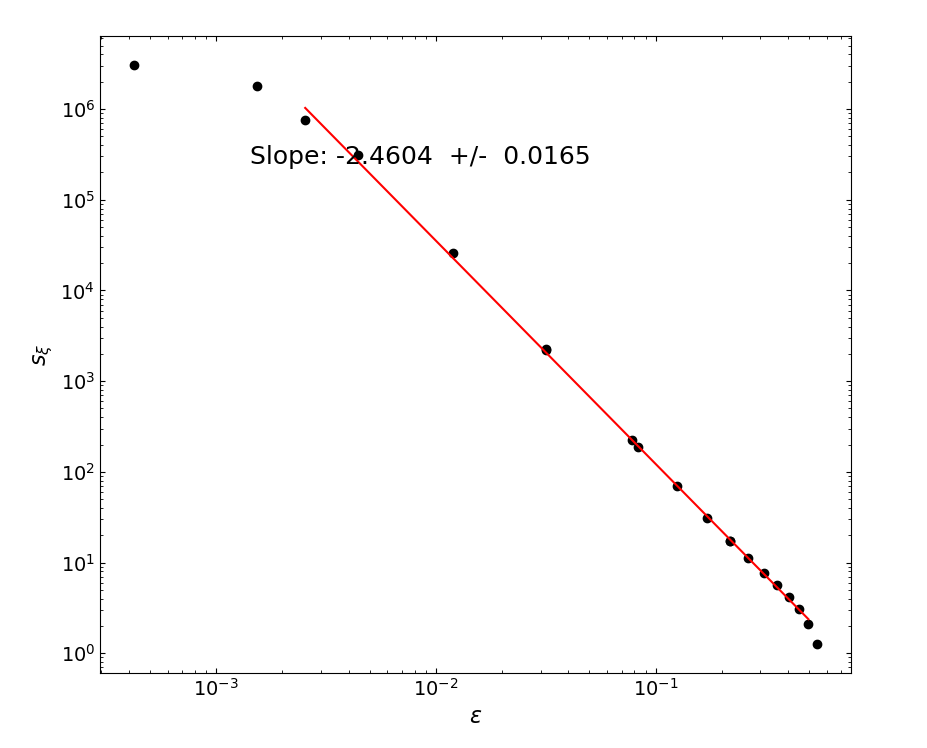}
	\caption{\footnotesize We show the LLS fit for burst size distribution exponent, $\sigma$. We fit $f(z)$ to an exponential to determine the decay constant, which corresponds to $s_\xi$. The log-log behavior of the cutoff burst size, $s_\xi$ as a function of $\epsilon = (T_c-T)^{1/\sigma}$ $1/\sigma$. We find $1/\sigma = 1/2.46 \approx 0.41 \pm .02$. }
	\label{fig:crit_sigma}
\end{figure}

With $\sigma$, one can work out the how the moments of size distribution $n_s$ behaves near the critical point, and derive relationships between the scaling of the generating function, average cluster size and order parameter in terms of $\tau$ and $\sigma$. These are:

\begin{equation}
2-\alpha = \frac{(\tau-1)}{\sigma}
\label{eq:zeroMoment}
\end{equation}

\begin{equation}
\beta = \frac{(\tau-1)-1}{\sigma}
\label{eq:betaRel}
\end{equation}

\begin{equation}
\gamma = \frac{2-(\tau-1)}{\sigma}
\label{eq:gammaRel}
\end{equation}

\begin{equation}
\nu = \frac{1}{\sigma D_s}
\label{eq:nuRel}
\end{equation}

We shall briefly go through these relations individually and establish the appropriate conclusions. We begin with the generating function, $G$. In practice, care is needed in determining the scaling of $G$, since the number of clusters is a function of the boundary/lattice constraints. In generating the clusters to determine $G$, one is naturally led to the relation $G \sim N_s/\left\langle S \right\rangle$ where $N_s$ is the total number of sites grown in the lattice, and $\left\langle S \right\rangle$ is the average cluster size. Because we derive the statistics for different $\epsilon_{T}$ keeping $N_s$ the same, this ultimately means that our measures for $G$ and $\left\langle S \right\rangle$ would be inversely proportional and lead to a faulty measure. Instead, we determine $G$ by keeping $N_s$ constant, but allow $N_s$ to vary when determining $\left\langle S \right\rangle$. This leads to scaling of $G$ with exponent $1-\alpha=0.9 \pm .1$.

For critical systems the zeroth moment is related to the scaling of the heat capacity for a typical thermal critical transition, where the specific heat capacity is the second derivative of the free energy,  $c_v \sim \partial_T^2 f \sim \epsilon^{-\alpha}$. As a result, the exponent is shifted by two such that $c_v \sim \partial_T^2 f \sim \epsilon^{-\alpha}$. In our case, because we work with the burst frequency distribution rather than the density distribution, normalized by burst size, $s$, we shift alpha by 1 rather than 2 in RP. Therefore, we report  $\alpha \approx 0.1$ which is somewhat curious. For RP, MF, and typical fluid or magnet thermal transitions give $\alpha = -2/3$, $\alpha=0$, and $\alpha=0-0.2$\cite{stanley1971phase}. There is some discrepancy reported in the literature with the percolation mean-field limit and the Ising mean-field limit, where $\alpha = -1,0$ respectively, and may result from differences in distributions used to derive values(as we showed).  

The zeroth moment is also found with the burst distributions exponents as shown in Eq.\ref{eq:zeroMoment} which is the typical relation in RP. However, the situation is different for AIP, and undermines the relation's applicability. When analytically solving the moment integrals we can represent the integrals as,
\begin{equation}
M_k =\epsilon^{\frac{1+k-\tau}{\sigma}} \int_{0}^{\infty}dz\ z^{k-\tau}f[z]
\label{eq:momentInt}
\end{equation}
where we must establish the convergence at the two limits. In the zero limit we can Taylor expand the integrand near zero which leads to $z^{1+k-\tau}f'[0]$. From this we can determine that in order for the integral to converge in the zero limit, we require $1+k-\tau>0$. Since $\tau\approx1.6$, this leads to the condition $k>0.6$. Thus for the zeroth moment the integral fails to converge at the lower limit. In the upper limit the integrand is exponentially suppressed by function $f[z]$. We therefore conclude that the relation \ref{eq:zeroMoment} doesn't hold for AIP. However, we can still determine the scaling of $\alpha$ as was done before. 

Next we consider among the most import quantities in characterizing the behavior of a phase transition, the order parameter. In RP we find that the order parameter is a function of the first moment and Eq\ref{eq:betaRel} establishes the relation between its exponents. We can check that the lower limit of integral converges since $k=1>0.6$. However, there exists a problem of a different nature. Namely, in RP we find that the order parameter is proportional to the site density. Near the critical point we observe that the site density becomes scale invariant and is described by $\rho_c \sim L^{d-D_f}$. However, for AIP regardless of the threshold, $T$, we find that the density is always described by $\rho_c$. That is, this quantity is determined exactly by fractal dimension scaling, $D_f$ and has no dependence on the control parameter. This undermines the existence of this type of order parameter. 

Despite this we can still attempt to interpret the order parameter to be the fraction of sites belonging to the largest burst. The resultant scaling near the critical point is found to be $P \sim \epsilon_{T}^{0.66}$ only very close to critical point, $\epsilon_{T}<10^{-2}$. Coincidentally, this fails to match the expected relation given by Eq.\ref{eq:betaRel} $(1.59-1)/0.41\approx1.46$. The reason being is that when analyzing Eq.\ref{eq:momentInt}, the upper limit contribution is dominated by $P_\infty=\int dz \ z^{1-\tau}f[0]$ which converges if $\tau>2$. This is a condition our $\tau$ fails to meet.

These inconsistencies highlight the key differences between RP and AIP models. In AIP, all bursts are stochastic realizations of RP's incipient infinite cluster. Certainly, if we looked at IP without any burst structure, there would only ever be a single cluster growing indefinitely, so an invaded site is certain to belong to the infinite cluster, and the wrapping probability is also one. Moreover, an order parameter does not seem appropriate in IP as it is not clear how one would define distinct phases such that at the critical point, IP is transitioning from one phase to the other through the critical formulation. Either the order parameter does not exist without further implementing behaviors to the model, or it is non obvious. 

The deficiencies of the previous two quantities do not plague the third, the mean cluster size $\left\langle S \right\rangle$. The integral conditions for convergence all exist for the second moment. We calculate the scaling of the second moment which is of same form as Eq.\ref{eq:gammaScaling} and leads to scaling $\epsilon_{T}^{-0.998 \pm .001}$. If we refer to Eq\ref{eq:gammaRel} we get $\frac{2 - 1.59}{0.41} = 1.0$ which is exactly what we expect, despite $\gamma \sim 1$ being very different from RP($\gamma_{RP}=43/18\approx2.39$). 


\begin{figure*}[t] 
	\centering
	\includegraphics[width=1\textwidth]{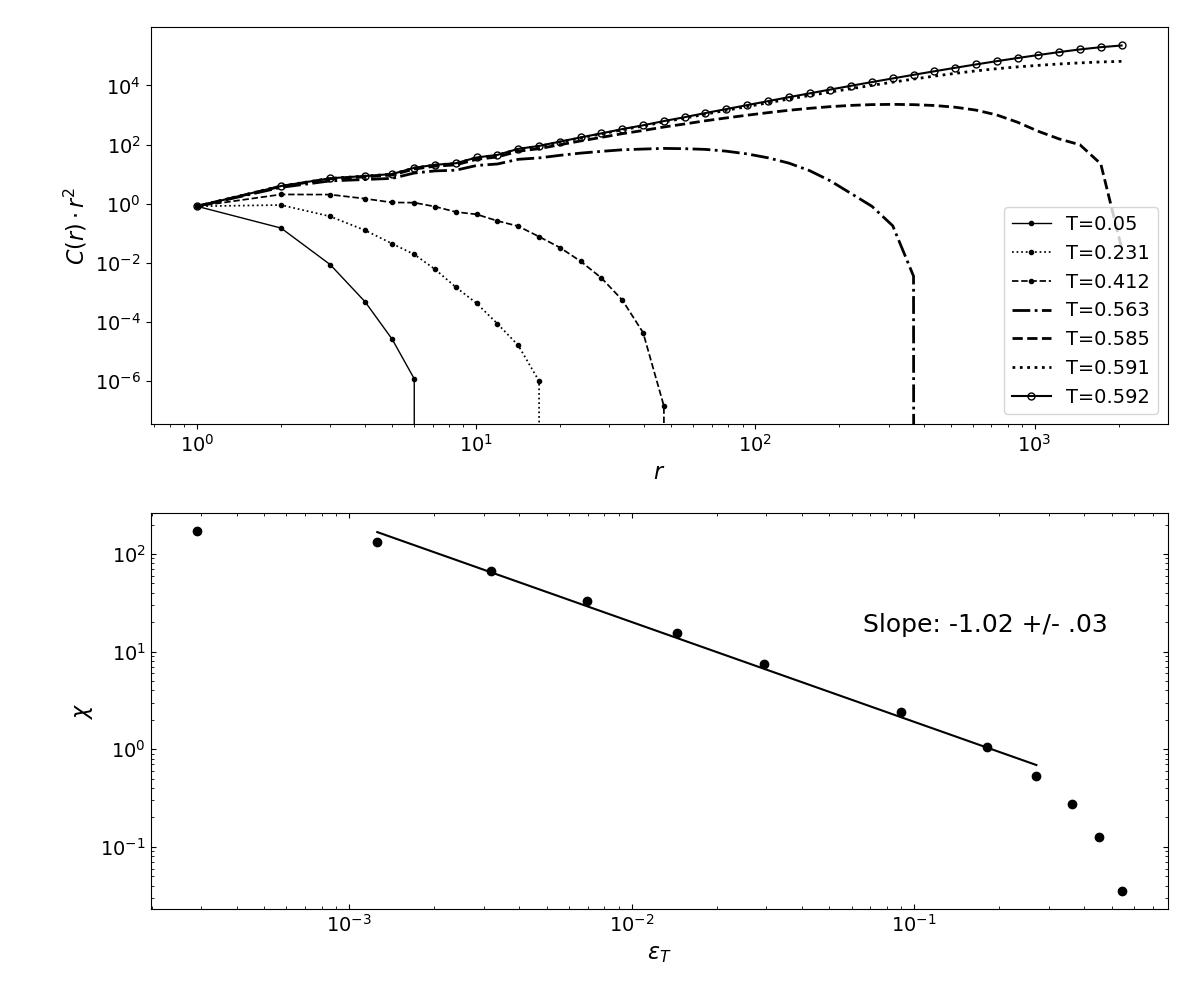}
	\caption{\footnotesize We show the results of th susceptibility divergence as a function of the pairwise correlation function. (Top) We show the behavior of the pairwise correlation function $C(r,T)$ as a function threshold, $T$ and distance between sites $r$. We observe the expected general behavior $C(r,T)\sim r^{2-\eta}f[-r/\xi(T)]$. (Bottom) We show the LLS fit of $\chi$ as a function of $\epsilon_{T}$ and find a scaling consistent with that of the mean burst size. Note deviations on small scale result from discrete lattice spacing. }
	\label{fig:crit_susc}
\end{figure*}

More importantly, in order to establish the consistency of results, we expect the susceptibility to diverge with the same exponent as the mean cluster size. We can calculate the susceptibility utilizing the fluctuation dissipation relation given by \cite{ma2018modern}, 
\begin{equation}
\chi \sim \frac{\beta}{V} \int_V d^dr \ C(r)
\label{eq:flucDiss}
\end{equation}
where $C(r)$ is the two-point correlation function as a function of distance between sites, $r$. Usually in thermal critical transitions, the susceptibility or the correlation function is itself a function of the order parameter fluctuations. In RP this also holds except there is an additional term corresponding to the mean cluster size. This leads to the following relation,
\begin{equation}
\chi = \left\langle S \right\rangle + \delta P^2
\label{eq:suscRel}
\end{equation} 
This offers yet another test for the order parameter as its fluctuations should contribute to the susceptibility divergence. In RP, both terms diverge with the same exponent\cite{coniglio1980fluctuations}. However, we find the order parameter fluctuations are only weakly consistent with the susceptibility since $\delta P^2 \sim \epsilon_{T}^{-0.89 \pm .05}$, when the order parameter is taken to be the fraction of sites belonging to the largest cluster (an obvious but dubious extension). When we evaluate the divergence of the susceptibility using Eq.\ref{eq:flucDiss} we get an exponent $\gamma=1.03\pm .05$ which is consistent with our mean burst size scaling, but beyond the error bounds of $\delta P^2$. The larger magnitude of the exponent of $\left\langle S \right\rangle $ as compared to $\delta P^2$ means that $\left\langle S \right\rangle$ will dominate the behavior near the divergence.

The divergence of the susceptibility as a result of the pair-wise correlation function nicely leads us into the discussion of the correlation length, $\xi$, which also happens to be the final critical exponent relation we aim to discuss from the list, Eq.3.17-20. In some sense it might be more reasonable to begin with $\xi$, since after all, the behavior of the correlation length is the fundamental ingredient to explaining the emergence of scale invariance as is posited by the finite-size scaling hypothesis. We independently determine the scaling of the correlation length in two ways. The first is done according to Eq.\ref{eq:corrLength}. For each burst of size $s$, we compute $R_s$ (Eq.\ref{eq:gyr_radius}) and compute the weighted average according to Eq.\ref{eq:corrLength}.  However, as with all the other exponents we find a scaling exponent, $\nu\approx1.3$ which is different from RP given the statistical error bounds. Table \ref{table:critValues} shows the values of the critical exponents. 

The second method utilizes the correlation function, $C(r)$, directly which describes the probability that two sites separated some distance, $r$, are likely to belong to the same burst. To calculate $C(r)$, we first grow an ensemble of clusters (100) within a lattice of size 4096x4096 with PEB . These clusters are broken up into bursts as a function of threshold, $T$. For each burst we calculate $C(r)$ according to standard counting method \cite{turcotte1997fractals}, and average over all bursts to generate an average $\left\langle C(r) \right\rangle_b $ which is then averaged across the ensemble to produce an average for a given lattice size, $\left\langle \left\langle C(r) \right\rangle  \right\rangle_L $. This is repeated for a range thresholds in order to establish the behavior in the regime of the critical value. We find that the behavior matches the expected behavior in 2d, $C(r)\sim r^{-\eta}e^{-r/\xi(T)}$, and based on the behavior very near the critical point we find $\eta=0.23 \pm .02$.

The top plot in Fig.\ref{fig:crit_susc} shows the effect of correlation length, $\xi$, which characterizes the length above which the likelihood of two sites to occupy the same burst becomes exponentially suppressed. As $T$ increases so too does the correlation length such that near the critical point the length approaches the system size. This motivates the general relation that near the critical point, the correlation length behaves according to $\xi \sim \epsilon_T^{-\nu}$ which we will subsequently discuss, but first we discuss how the susceptibility diverges with the correlation length. Following Eq.\ref{eq:flucDiss}, we can compute the integral in 2d using, 
\begin{equation}
\begin{split}
\chi &\sim \int_{0}^{\xi}dr \langle\langle C(r) \rangle\rangle_L \\
     &= \epsilon_{T}^{-\nu(1-\eta)} 
\end{split}
\end{equation}
In the previous section we found exponent $\nu\approx 1.3$ using the typical RP definition of $\xi$ defined by Eq.\ref{eq:corrLength}. From the relation above we can check that $\gamma \approx 1.3(1-0.23) \approx 1.0 $. 

We can also assess the behavior of the correlation length by counting the number of sites contained in the largest burst. Since the density of sites is fixed the largest burst, $s_\infty$, should scale according to the correlation volume, $\xi^2 = \epsilon_{T}^{-2\nu}$. From LLS fit of $s_\infty$ vs $\epsilon_{T}$ we find $\xi \sim \epsilon_{T}^{-1.3 \pm .03}$ which is within error bounds of our previous determination of $\nu$.

As a final check we can utilize the burst size scaling with radius of gyrations given by $s\sim R_s^{D_s}$ to check relation \ref{eq:nuRel}. We find by LLS fit that $D_s = 1.859 \pm .002$ and as shown in the previous section $\sigma=0.41 \pm .02$. This gives a value of $\nu= 1.31$ which is consistent within error bounds of the other results which all yield values in the vicinity of $1.3$. 

As a brief conclusion to this section, we found extrapolating from RP's behavior, we would expect mean square fluctuations in the order parameter coupled with the divergence of the correlation length to give rise to observed cluster structure of the model. However, we quickly ran into problems with the order parameter both in conceptually defining the notion of distinct phases and formally deriving appropriate consistency relations between relevant quantities. Therefore, there was no consistent characterization of the order parameter, but we also found in characterizing the correlation length and mean cluster size that we were able to account for the critical divergence of susceptibility with these quantities alone. In RP the emergence of macroscopic connectivity is nicely characterized by the divergence of order parameter fluctuations, and its a particularly nice conceptual feature that order parameter fluctuations and random site fluctuations should scale similarly. However, it does not seem necessary in describing the critical behavior of AIP( and perhaps more broadly applies to self-organizing critical behavior). What IP simulates is a growth mechanism, and if we introduce a threshold formulation to define distinct bursts we find that we can define regimes of critical growth or subcritical growth depending on the threshold, $T$. This characterization better fits within the description of metastable nucleation of critical droplets. 

\section{Hyperscaling}
We continue to delve into the critical behavior of AIP by analyzing whether or not hyperscaling is a feature of the model. While not an essential prerequisite for criticality, its presence does highlight the existence of additional relations which simplify the quantities needed to explain the observed behavior. Generally, systems with hyperscaling can be characterized by independently determining only 2 critical exponents. All others can be determined by the various relations including the additional hyperscaling relations. Still, compared to other notions we have discussed, the concept of hyperscaling is generally more loosely defined and essentially amounts to the criteria where there exist relations between exponents contain the dimensionality of the system. As an example, we will illustrate among the more insightful of these relations and how it manifests itself in the system. 

Perahps the most significant quantity is the correlation length of the system, $\xi$. With $\xi$ we can determine the likelihood two sites are to be related to one another. In RP one would expect that this should be related to whether or not the two sites belong to the same cluster, and in RP this statistically works out to be the case. We can see this with the following calculation. We consider a box with sides of length $\xi$. The number of clusters contained within the box should remain constant with threshold. In particular, near the critical point where $\xi$ increases (and the size of our correlation box), the clusters contained should scale proportionally such that number of clusters does not change.

To carry out this calculation, we can make use of $G$ which is the zeroth moment of the cluster distribution and amounts to the number of clusters per site or the cluster density. If this quantity is multiplied by $\xi^2$ we should get the number clusters in a correlation volume given by, 
\begin{equation}
N_\xi = G \ \xi^2 \sim \epsilon^{2-\alpha - 2\nu} \sim constant
\end{equation}
For RP $\alpha=-2/3$ and $\nu=-4/3$ yields an average cluster count which is constant. This is indeed what one would expect if the correlation length represents the spatial extent of observed clusters. Note, also that this leads to the following hyperscaling relation,
\begin{equation}
2-\alpha = d\nu
\label{eq:hyperscaling}
\end{equation}
Because we can also determine $2- \alpha$ from the cluster distribution exponents, this also means that we can determine the scaling of the correlation length based entirely on the dimensionality of the system, a very powerful result. Similarly, all other relations will be the result of the two exponents $\tau$ and $\sigma$

Much of the same formalism can be applied directly to our AIP model, where we have characterized our own values for the critical exponents. First we define the appropriate critical parameter $\epsilon_{T}$ as was shown in section \ref{section:critThresh},

\begin{equation}
\epsilon_{T} = \frac{T-T_c}{Tc}
\end{equation}
where $T$ is the burst threshold and $T_c$ is the critical burst threshold. The critical values of the scaling exponents are reported in Table \ref{table:critValues}. However because $n_s$ for us is a frequency distribution rather than a density distribution, we shift the exponent by 1. If we similarly calculate the number of bursts in correlation volume,
\begin{equation}
N_\xi^{\prime} = G \xi^2 \sim \epsilon_{T}^{-0.6}
\label{eq:numExcess}
\end{equation}
we find a scaling that diverges with value $\approx -0.7$ in contrast with the RP case which remained constant. 

This is another key differences between the apparent criticality present in this AIP model and RP, which suggests that because the bursts are not realizations of order parameter fluctuations as in RP, we do not observe similar behavior. Instead, there appears to be an excess of bursts contained within a correlation volume and scales according to $N_{\xi}^{\prime}$. This observation introduces a new puzzle, but also establish the lack the hypersaling in our model, namely Eq.\ref{eq:hyperscaling}. The absence of hyperscaling in IP models was also found by \cite{willemsen1984investigation}. The authors found that the cluster distribution exponent $\tau$ could not be determined by spatial dimension $d$ according to hyperscaling relation, $\tau = 1+d/D_f$.

Further evidence comes from \cite{ray1988crossover} which also observe the absence of hyperscaling in the mean-field regime of long-range bond percolation. They observe differences between scaling exponents, $D_f$ and $D_s$(as defined here) and suggest that there are differences cluster density for small versus large clusters. We can add to this line of inquiry by similarly noting a difference between exponents, $D_f$ and $D_s$.(Though the differences are small, the differences are outside error bounds. Also, both measures are characterizations of many scales and consequently are rather insensitive measures, so any observed difference can have quite notable consequences) One possible explanation is that this suggests that their might be two competing correlation mechanisms. We can think about the density of sites versus the density of clusters. One correlation mechanism that is inherited from RP and is present in the IIC of the original IP cluster which would characterize the density of sites, given by $d-D_f$. The second correlation length comes about when we introduce the notion of bursts, and burst density would therefore be characterized by $2-D_s$. That these two are not the same is to say that site and burst density do not follow the same scaling.  

\begin{figure*}[t] 
	\centering
	\includegraphics[width=1\textwidth]{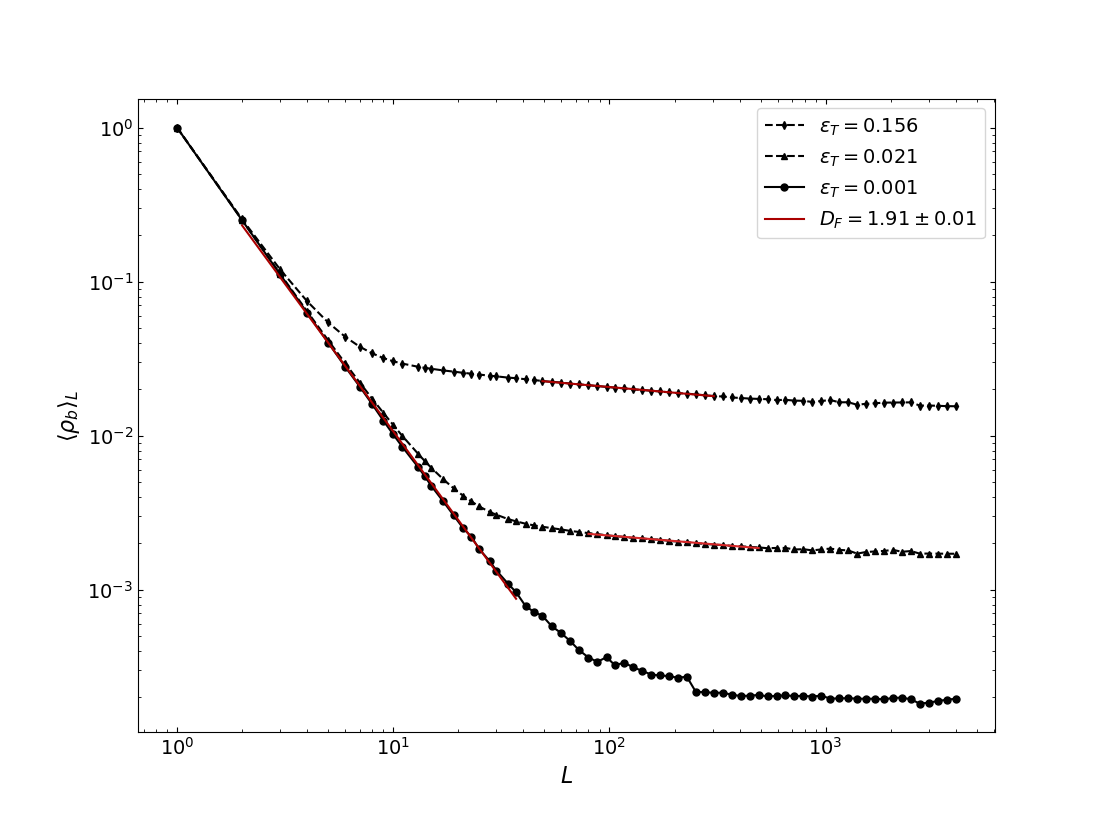}
	\caption{\footnotesize We show the results of the crossover behavior inherent to average burst epicenter density, $\langle \rho_b \rangle_L$ as a function of length scale, $L$. The correlation length defines the scale where we expect power-law behavior where the average number of burst,$\langle N(L)\rangle$ in region size, $L^2$ scales according to the expected mass scaling of individual sites(as is expected from scale invariance), $N(L) \sim L^{D_f}$. on larger length scales, the scaling becomes uniform.    }
	\label{fig:epicenter_crossover}
\end{figure*}

We briefly provide evidence for this by computing the average density of burst epicenters as function of length scale, $l$. For each burst we compute the epicenter as the burst center of mass. For each length scale, we can easily compute the average density according $\langle \rho_b \rangle_l =\langle N_b\rangle/l^2$, where $N_b$ is the number of burst contained in the box element, $l^2$. We determine the correlation length by extrapolating where the power-law behavior intersects the homogeneous scaling regime. We repeat this process for multiple values of critical scaling parameter $\epsilon_T$ to derive the correlation scaling for epicenters, $\xi_b \sim \epsilon_{T}^{-\nu_b}$ where $\nu_b$ defines the burst epicenter scaling. We find $\nu_b \approx 0.5$ which is the mean-field value, and the expected value if burst epicenters were distributed essentially according to a random walk. For length scales above $\xi_b$(nearly flat regions in upper curves of Fig\ref{fig:epicenter_crossover}) we observe scaling given $d-D=0.127$, which yields a result consistent with $D_s$ scaling.    

We can also refer to \cite{klein2007structure} which built upon previous findings to understand a similar discrepancy when analyzing the behavior mean field clusters in the Potts 2d model. They found they could account for the excess of clusters in a correlation volume if instead cluster density was the result of random walk distribution of clusters, and developed the notion of "fundamental clusters" which do have the property that their density matches that of site density. In their analysis they found that the additional random walk element would contribute to number scaling by a factor of $\epsilon^{1/2}$, since the resulting spatial distribution is Gaussian whose width is given by $\sigma = \sqrt{N}$ where $N$ scales as $G \sim \epsilon^{1}$. This is nearly of the same magnitude needed to account for our own excess. 

\section{Discussion}
We find plenty of examples in the literature suggesting a connection between critical phenomena and seismicity. Naturally, one might ask: must we insist on the presence of a phase transition in order to describe the underlying mechanisms of seismicity? There are a number of challenges Our AIP is an example of We found that our AIP model, a characteristic example of self-organized criticality (SOC), has many of the essential features of criticality, except it does not inherit any phase transition mechanics(namely, no distinct phases and no notion of an order parameter). These features arise because our model preserves a  critical point where the correlation length of bursts diverges. This leads to the conclusion that in systems which are already described by the absence of a characteristic length scale, the transition from a finite to infinite correlation length need not describe a state with finite correlation length (which may never have existed) to a state described by an infinite correlation length. 

The absence of a phase transition is further supported because in AIP the site density remains constant, and is described as a feature of static network properties as shown in section \ref{section:boundaryConditions}. These static features arise independent to the pseudo-critical mechanisms we define here and in the previous sections, and are therefore insensitive to the tuning of the critical parameter, $T$. This fundamentally leads to absence of hyperscaling, where the absence of hyperscaling indicates the existence of multiple correlation lengths in the system, further undermining the applicability of phase transition mechanics. This is in contrast to RP, where the phase transition is described by a mapping between the site density fluctuations and the observed cluster fluctuations, describing the emergence of global connectivity. This is understood from the observation that the pair distribution function is isomorphic to the pair connectedness function \cite{klein2007structure}.

Nevertheless, the absence of the order parameter and its critical fluctuations does not prevent the system from exhibiting scale invariant structure. In fact, the situation is similar to the conditions present during seismicity where the system seems imbued with an infinite correlation length as is described by \cite{sornette1989self}, and subsequent behavior exists atop underlying long-range correlated behavior. Therefore, rather than describing a phase transition we have only a stochastic growth mechanism with a notion of independent realizations that gives rise to a critical type burst distribution. The characterization of this distribution and the correlation length is adequate in defining the universality class of the model's inherent randomness and the growth mechanism by which the fluctuations grow to all scales. 


One of the claimed benefits of SOC models is that they do not require a tuning parameter, our AIP is characterized by a critical distribution similar to Fisher's critical droplet distribution or that near the Ising mean-field spinodal. Thus, in order to produce scale invariant bursts(even atop the underlying fractal substrate) the control parameter must be near its critical value. However, it is possible to rescue the idea that AIP does not explicitly need to be fine tuned to be near its critical and reclaim one of the primary advantages believed to be a general feature of SOC. As \cite{gabrielli2000invasion} stress, nearly all SOC systems have some notion of a driving parameter, usually a ratio which characterizes the growth of the system. As was shown in Section \ref{section:critThresh} Eq.\ref{eq:B2Bratio}, we can recast the threshold as a bulk-to-boundary ratio which serves as the driving parameter describing burst growth. A system's equilibrium growth mechanics might then tend to the critical ratio.

In conclusion, the AIP model is a hybrid model with a Fisher type burst distribution characterized with near mean-field exponents $\tau_{AIP}=2.59$, $\sigma_{AIP}=0.41$. Again, the normalized per site mean-field cluster distribution is characterized with exponents $\tau_{MF}=5/2$ and $\sigma_{MF}=1/2$ respectively. This naturally implies that AIP is similar to but distinct from mean-field and ought to belong to its own universality class, since the exponents are all different from mean field as shown in Table\ref{table:critValues}. Many of the static features are inherited from RP's IIC, and the addition of the avalanche-burst mechanism introduces a correlation length for bursts that is consistent with mean-field, since $\nu \sim 1/2$. Despite the presence of multiple correlation lengths, the system maintains its fractal properties and is largely self-similar on all length scales. While the finite size scaling hypothesis does lead to manifestly scale invariant systems, in practice it is likely a constraint that need not be enforced, and instead more investigations should seek to understand how multiple correlation lengths can effectively preserve the fractal properties which dominate so many complex systems. 

Finally, because we inherit the long-range correlations present during the critical phase of RP, its fair to wonder how the behavior changes with different underlying correlation structure. This will be the subject of next chapter where we consider the presence of long-range correlations present between sites described by $C(r) \sim r^{-H}$, where $r$ is the distance between sites. This should yield distinct critical distributions, $n_s(T,\sigma, H)$, and thereby lead to a potential family of distributions with distinct burst scaling behavior. This is what we will seek to characterize in future work.

\section{Acknowledgements}
We would like to thank Bill Klein for his guidance and his enlightening and stimulating conversations. The research of RAO and JBR has been supported by a grant from the US Department of Energy to the University of California, Davis. DOE Grant No. DE-SC0017324.

\bibliography{AIP_bibliography}
\end{document}